\documentclass[11pt,letter]{article}
\usepackage{algorithm}
\usepackage[noend]{algorithmic}
\usepackage{epsfig}
\usepackage{graphicx,geometry}
\usepackage{amssymb,amsfonts,amsbsy,latexsym}
\usepackage[small,compact]{titlesec}

\newtheorem{theorem}{Theorem}[section]
\newtheorem{lemma}[theorem]{Lemma}

\newtheorem{corollary}[theorem]{Corollary}
\newtheorem{observation}[theorem]{Observation}

\newenvironment{proof}[1][Proof]{\begin{trivlist}
\item[\hskip \labelsep {\bfseries #1}]}{\end{trivlist}}
\newcommand{\qed}{\ensuremath{\square}}

\renewenvironment{proof}
{\noindent\textbf{Proof:\ }} {\hspace*{\fill}\qed}

\newenvironment{sketch}
{\noindent\textbf{Sketch of the proof:\ }} {\hspace*{\fill}\qed }

\author {
Petra Berenbrink \thanks {\small School of Computing Science, Simon Fraser University, Burnaby, BC, V5A 1S6, Canada }
\and
Colin Cooper \thanks {\small Department of Computer Science, King's College, London, WC2R 2LS, UK }
\and
Zengjian Hu  \thanks {\small School of Computing Science, Simon Fraser University, Burnaby, V5A 1S6, Canada}
}

\title{Energy Efficient Randomised Communication in Unknown AdHoc Networks
}

\begin{document}

\maketitle

\begin {abstract}
This paper studies broadcasting and gossiping algorithms in 
random and general AdHoc networks. Our goal is not only to minimise the broadcasting and gossiping time, but also to minimise the {\em energy consumption}, which               
is measured in terms of the total {\em number of messages} (or {\em                    
transmissions}) sent.                                                                       
We assume that the nodes of the network do not know the network, and that they can only send with a fixed                   
power, meaning they can not adjust the areas sizes that their messages                 
cover. We believe that under these circumstances the number of transmissions is              
a very good measure for the overall energy consumption.                                
																					   
For random networks, we present a broadcasting algorithm where every node              
transmits at most once. We show that our algorithm broadcasts in $O(\log               
n)$ steps, w.h.p, where $n$ is the number of nodes. We then present a                   
$O(d \log n)$ ($d$ is the expected degree) gossiping algorithm using $O(\log n)$ messages per node.                                

For general networks with known diameter $D$, we present a                             
randomised broadcasting algorithm with optimal broadcasting time $O(D \log             
(n/D) + \log^2 n)$ that uses an expected number of $O(\log^2 n / \log                   
(n/D))$ transmissions per node. We also show a tradeoff result between the broadcasting time and the                   
number of transmissions: we construct a network such that any oblivious algorithm             
using a time-invariant distribution requires $\Omega(\log^2                             
n / \log (n/D))$ messages per node in order to finish broadcasting in                  
optimal time. This demonstrates the tightness of our                                   
upper bound. We also show that no oblivious algorithm can complete               
broadcasting w.h.p. using $o(\log n)$ messages per node.
\end {abstract}

\section {Introduction}

In this paper we study two fundamental network communication
problems, {\em broadcasting} and {\em gossiping} in unknown AdHoc
networks. In an unknown network the nodes do not know their 
neighbourhood or the whole the network structure, the only the size 
of the network. The nodes model mobile devices equipped with
antennas. Each device $d$ has a fixed {\em communication range}, meaning that
it can listen to all messages send from nodes within that range, and all nodes in that range can receive messages from $d$.  We do not assume that $d$ can sent with with different power levels, hence the communication range is fixed. Note that we allow different communication ranges for different nodes. If
several nodes $d$'s communication range send a message at the same
time, these messages {\em collide}, the device is not able to
receive any of them.  Note that a node does not know which
nodes are able to the receive messages it sends, and the node might
not know all neighbours in his own communication range.  Since the
communication ranges of different devices can vary, one device may
be able to listen to messages send out by a node in its
communication range, but not vice-versa. This forbids the
acknowledgement-based protocols since the receiver might not be able
to send a confirmation message to the sender. Another challenge in
these networks is that, due to the mobility of the nodes, the
network topology changes over time. This last characteristic makes
it desirable that communication algorithms use local information
only. Mobile devices tend to be small and have only small
batteries. hence, another important design issue for communication in ad-hoc
networks is the energy efficiency (see, e.g., \cite {HKB99,MFHH02,KKKP97}) of protocols.

In this paper we design efficient communication algorithms which
minimise the broadcasting or gossiping time, and which also minimise the energy
consumption. We measure the energy consumption in terms of the number of total transmissions. We believe that the number of transmissions is              
a very good measure for the overall energy consumption since  
we do not assume variable communication ranges. We also show that
there is a trade-off between minimising the broadcast or gossiping
time, and the number of messages that are needed by randomised
protocols.

The rest of the paper is organised as follows.  The rest of this
section introduces the related work, our model, and our new results.
Section \ref {se:rand-broadcast} and Section \ref {se:rand-gossip}
study broadcasting and gossiping for random networks. In Section
\ref {se:arbitrary}, we analyse an broadcasting on general (not
random but fixed) networks with known diameter. Our algorithm
minimises both the broadcasting time and the number of
transmissions. Finally, in Section \ref {se:lb-arbitrary} we show
some lower bounds on broadcasting time and the number of used
messages.

\subsection {Related Work} \label {se:related}

Here we only consider randomised broadcasting and gossiping protocols
for unknown AdHoc networks. For an overview of deterministic
approaches see \cite{KP03}. Let $D$ be the diameter of the network.
\paragraph {Broadcasting}
Alon et. al \cite {ABLP91} show that there exists a network with
diameter $O(1)$ for which broadcasting takes expected time $\Omega
(\log ^2 n)$.  Kushilevitz et. al \cite{KM98} show a lower bound of
$\Omega (D \log (n/D))$ time for any randomised broadcast algorithm.
Bar-Yehuda et. al \cite {BGI92} design an almost optimal
broadcasting algorithm which achieves the broadcasting time of $O((D
+ \log n) \log n)$, w.h.p.. 
Later, Czumaj et. al \cite {CR06} propose an elegant algorithm which
achieves (w.h.p.) linear broadcasting time on arbitrary networks.  Their
algorithm uses carefully defined selection sequences which specify the
probabilities that are used by the nodes to determine if they will sent a
message out or not. This algorithm needs $\Theta (n)$ transmissions per
node. Czumaj et. al \cite {CR06} also obtain an algorithm under the
assumption that the network diameter is known. The algorithm finishes
broadcasting in $O(D \log (n/D) + \log^2 n)$ rounds, w.h.p, and uses
expected $\Theta (D)$ transmissions per node. Also, independently,
Kowalski et.al \cite {KP03} obtain a similar randomised algorithm with the
same running time.

Els{\"a}sser and Gasieniec \cite {EG05} are the first to study the
broadcasting problem on the class of directed random graphs
${\mathbb G}(n,p)$.  In these networks, every pair of nodes is
connected with probability $p$. They propose a randomised algorithm
which achieves w.h.p. strict logarithmic broadcasting time. Their
algorithm works in three phases: In the first phase (containing
$D-1$ rounds), every informed node transmits with probability 1. In
the second phase, every informed node transmits with probability
$n/d^D$, where $d = np$ is the expected average degree of the graph. 
In the third phase, every node informed in the first two
phases transmits with probability $1/d$.

In \cite {Els06}, Els{\"a}sser studies the communication complexity of 
broadcasting in random graphs under the so-called {\em random phone call} 
model, 
in which every user forwards its message to a randomly chosen neighbour at 
every time step. They propose an algorithm that can complete 
broadcasting in $O(\log n)$ steps by using at most  
$O(n \max \{ \log \log n, \log n / \log d \})$ transmissions, which is optimal
under their random phone call model.

\paragraph {Gossiping}

For gossiping, all the previous works follows the join model, 
where nodes are allowed to 
join messages originated from different nodes together to 
one large message.
 So far the fastest randomised algorithm for
arbitrary networks has a running time of $O(n \log ^2 n)$ \cite {CR06}. The
algorithm combines the linear time broadcasting algorithm of \cite
{CR06}, and a framework proposed by \cite {CGR01}. The framework
applies a series of limited broadcasting phases (with broadcasting
time $O(f(n))\,$) to do gossiping in time $O(\max \{ n \log n, f(n)
\log ^2 n) \}$.
Chlebus et. al \cite {CKR06} study the average-time complexity 
of gossiping in Radio 
networks. They give a gossiping protocol that works in average 
time of $O(n/\log n)$, which is shown to be optimal. For the case when 
$k$ different nodes initiate broadcasting (note that it is gossiping when $k=n$), they give an
algorithm with $O(\min \{k \log (n/k) + n / \log n \})$ 
average running time.

\paragraph {Random Graphs}
In the classic random graph model of Erd{\"o}s and Reny{\'i},
${\mathbb G} (n,p)$ is a $n$-node graph where any pair of vertices is
connected (i.~e.~, an edge is built in between) with probability $p$.
It can be shown by Chernoff that every node in the network has $\Theta
(d)$ neighbours w.h.p. Moreover, It is well known (see e.g. \cite
{B90}, \cite {CL01}) that as long as $p = \Omega (\log n / n)$, the
diameter of the graph is $(1 + o(1))(\log n / \log d)$ w.h.p.
Besides,  if $p > \log n / n$, the graph is connected w.h.p.

\subsection{The Model} \label {se:model}

We model a radio network is modeled by a directed graph $G = (V,E)$.
$V$ is the set of mobile devices and $|V| = n$. For $u,v\in V$,
$(u,v) \in E$ means that $u$ is in the communication range of $v$
(but not necessarily vice versa). We assume that the network $G$ is
unknown, meaning that the nodes do not have any knowledge about the
nodes that can receive their messages, nor the number of nodes from
which they can receive messages by themselves.  This assumption is
helpful since in a lot of applications the graph $G$ is not fixed
because the mobile agents can move around (which will results in a
changing communication structure). 

We assume that $G$ is either arbitrary \cite {ABLP91,CR06,KM98}, or
that it belongs to the random network class \cite {EG05}. For random
graphs, we use a directed version of the standard model ${\mathbb G}
(n,p)$, where node $v$ has an edge to node $w$ with probability $p$.
Let $d$ be the average in and out
degree of $G$. Recall that $d = np$ and $D= (1+o(1)) (\log n / \log d)$.

In the broadcasting problem one node of the network tries to send
a message to all other nodes in the network, whereas in the case of 
gossiping every
node of the network tries to sends a message to every other node.
The {\em broadcasting time} (or the {\em gossiping time}) denotes
the number of communication rounds needed to finish broadcasting (or
gossiping).  The {\em energy consumption} is measured in terms of
the total (expected) number of transmissions, or the maximum number
of transmissions per node.

\subsection {New Results} \label {se:newresults}

The algorithms we consider are {\em oblivious}, i.~e.~all nodes have to use the same algorithm.

\paragraph {Broadcast in random networks} Our
broadcasting algorithm is  similar to the one of Els{\"a}sser
and Gasieniec in \cite {EG05}. The difference is that our algorithm sends at most one message per node, whereas the randomised algorithm of \cite {EG05} sends up to  $D-1$ messages per node. The broadcasting time of both algorithms is $O(\log n)$, w.h.p. Our proof is very different from the one in
\cite {EG05}. Els{\"a}sser and Gasieniec show first some structural
properties of random graph which they then use to analyse their algorithm.
We directly bound the number of nodes which received the message after
every round. Our results are also more general in the sense that we 
only need $p = \omega (\log n / n)$ instead of 
$p = \omega (\log^{\delta} n / n)$ for constant $\delta > 1$ (see \cite {EG05}).

\paragraph {Gossiping in Random Networks} We modify the algorithm of \cite{CR06} and achieve a gossiping algorithm with running time $O(d \log n)$, w.h.p, where every node sends only $O(\log n)$ messages. To our best knowledge, this is the first gossiping algorithm specialised on random networks. So far, the fastest
gossiping algorithm for general network achieves $O(n \log ^2 n)$
running time and uses an expected number of $O(n \log n)$ transmissions per node
\cite {CR06}.

\paragraph {Broadcasting in General networks} Our 
randomised broadcasting algorithm for general networks completes 
broadcasting time $O(D
\log (n/D) + \log ^2 n)$, w.h.p. It uses an expected number of
$O(\log ^2 n / \log(n/D))$  transmissions per node. Czumaj and Rytter
(\cite{CR06}) propose a randomised algorithm with $ O(\log^2 n + D
\log (n/D))$ broadcasting time.  Their algorithm can easily be
transformed into an algorithm with the same runtime bounds and an
expected number of $\Omega(\log ^2 n)$ transmissions per node.

\paragraph {Lower Bounds for General networks} First we show a lower
bound of $n \log n /2$ transmissions for any randomised broadcasting
algorithm with a success probability of at least $1 -n^{-1}$. We assume that every node in the network uses the
same probability distribution to determine if it sends a
message or not. Furthermore, we assume that the distribution does
not change over time. To our best knowledge, all distributions used
so far had these properties.  Czumaj and Rytter (\cite{CR06})
propose an algorithm that needs $O(n\log^2 n)$ messages (see Section
\ref {se:related}). Hence, there is still a factor of $\log n$
messages left betwen upper and our lower bound.

Finally, using the same lower bound model, we show that there is a
network with $O(n)$ nodes and diameter $D$, such that every
randomised broadcast algorithm requires an expected number of at
least $\log^2 n /(\max\{4c, 8\} \log (n/D))$ transmissions per node
in order to finish broadcasting in time $ c D \log (n/D)$ rounds
with probability at least $1- n^{-1}$. This lower bound shows the
optimality of our proposed broadcasting algorithm (Algorithm \ref
{alg:arbitrary}).

\section{Broadcasting in Random Networks} \label {se:rand-broadcast}
In this section we present our broadcasting algorithm for random networks.
Our algorithm is based on the algorithm proposed in \cite{EG05}.   The
algorithm completes broadcasting in $O(\log n)$ rounds w.h.p, which
matches the result in \cite {EG05}.

Let $T = \left \lfloor \log n/\log d \right \rfloor$. Throughout the
analysis, we always assume that $n = |V|$ is sufficiently large, and
$p > \delta \log n/n$ for a sufficiently large constant $\delta$. Note
that the later condition is necessary for the network to be
connected. In the following, every node that already got the message
is called {\em informed}.
An informed node $v$ can be in one of two different states. $v$ is called
{\em active} as soon as it is informed, and it will become
{\em passive} (meaning it will never transmit a message again) as soon as it tried once to send the message.
\begin{algorithm}
\caption{An Energy efficient algorithm for Random Networks} \label{alg:broadcast}

\medskip
{\bf Phase 1:}

\begin{algorithmic}[1]
\STATE The state of the source is set to {\em active}.
\FOR {round $r = 1$ to $T$ }
    \STATE Every {\em active} node $v$ transmits once and becomes {\em passive}.
    \IF {node $v$ receives the message {\em for the first time}}
        \STATE {The status of $v$ is set to {\em active}.}
    \ENDIF
\ENDFOR
\end{algorithmic}
\medskip
{\bf Phase 2:}
\begin{algorithmic}[1]
    \IF {$p \le n^{-2/5}$}
        \STATE Every active node transmits with probability $1/(d^T p)$ and becomes {\em passive}.
        \IF {node $v$ receives the message {\em for the first time}}
            \STATE The status of $v$ is set to {\em active}.
        \ENDIF
    \ENDIF
\end{algorithmic}
\medskip
{\bf Phase 3:}
\begin{algorithmic}[1]
\FOR {round $r = 0$ to $\beta \log n$ ($\beta$ is a constant)}
    \IF {$p \le n^{-2/5}$}
        \STATE Every active node transmits with probability $1/d$
        \STATE A node that has transmitted  the message becomes {\em passive}.
    \ELSE
    \STATE Every active node transmits with probability $1/dp$
        \STATE A node that has transmitted  the message becomes {\em passive}.
        
    \ENDIF
\ENDFOR
\end{algorithmic}
\end{algorithm}


\noindent
The main idea of the algorithm is as follows.

\begin {enumerate} \itemsep 0pt
\item [{\bf Phase 1.}] The goal of Phase 1
 is to inform $\Theta \left (d^{T}\right )$ nodes w.h.p. (Lemma \ref {le:phase1-2}).
To prove this result, we repeatedly use Lemma \ref {le:phase1-1}, which bounds
the number of active nodes after each round.

\item[{\bf Phase 2.}] The goal of Phase 2 is to inform $\Theta(n)$ nodes w.h.p.
when $p \le n^{-2/5}$ (Lemma \ref {le:phase2}).
For the rest case we do not need Phase 2.

\item[{\bf Phase 3.}]
The goal of Phase 3 is to inform every remaining uninformed node in 
the network w.h.p. (Lemma \ref {le:phase3}).

\end {enumerate}

We prove the following theorem.

\begin {theorem} \label{thm:main}
If $p > \delta \log n /n$ for a sufficiently large constant $\delta$,
Algorithm \ref {alg:broadcast} completes broadcasting in $O(\log n)$ rounds, w.h.p.
Furthermore, every node performs at most one transmission
and the expected total number of transmissions is $O(\log n /p)$.
\end {theorem}


The number of transmissions performed in Phase 1 is
$1 + d + \ldots + d^{T-1} = O(1/p)$ since $T = \lfloor \log n / \log d \rfloor$.
The (expected) number of transmissions in each round of Phase 2 and 3 is bounded by $1/p$. Hence, the expected total number of transmissions is $O(\log n /p)$.

To proof Theorem \ref{thm:main} it remains to bound the broadcasting time. This part of the proof is split into several lemmata. Let $U_t$ be the set of
active nodes at the beginning of Round $t$, $Q_t$ be the set of
nodes which transmit in Round $t$. Let $N_t$ be the number
of not informed nodes at the beginning of Round $t$. We first prove
the following simple observations which will be used in the later
sections.

\begin {observation} \label {obv:relation}
~
\begin {enumerate} \itemsep 0pt
\item $\forall t \in [1,T]$, $U_t = Q_t$.
\item $\forall t \in [1,T], N_t = n - \left(\sum_{i=1}^{t-1} {|Q_i|} + |U_t|
\right).$
\item $\forall r,t \ge 1$, $r < t$, $|U_t| \ge |U_r| - \sum_{i=r}^{t-1} {|Q_i|}$.
\item $Q_i \bigcap Q_j = \phi$ for all $ i,j \ge 1$ with $i \ne j$.
\end {enumerate}
\end {observation}
\begin {proof}
$(1)$ is true since in Phase 1 of our algorithm 
every active node transmits. To prove $(2)$,
note that for any informed node $v$ at Round $t$, there are only two
possibilities: Either $v$ transmits in some round between $1$ and
$t-1$ (i.~e.~, $v \in Q_i, i \in [1,t-1]$), or $v$ must be active at
Round $t$, (i.~e.~, $v \in U_t$). For $(3)$, simply note that nodes
being active in Round $r$ will remain active until Round $t$ if they
do not transmit in the meantime. For $(4)$, note that every node
only transmits at most once per broadcast.
\end {proof}

Observation \ref {obv:relation}(4) helps us to argue that
the random experiments used later in the analysis are independent from
each other. In the following, we first prove Lemma \ref{le:phase1-1}
(1) showing that in each round of Phase 1 the number of active nodes
grows by a factor of $\Theta (d)$, w.h.p. The second part of Lemma
\ref{le:phase1-1} strengthen the results if the number of active
nodes is between $[\log^3 n, \frac{1}{p \log n}]$.

\subsection {Analysis of Phase 1}

\begin {lemma} \label {le:phase1-1}
If $p > \delta \log n /n $ and $1\le t\le T$ (Phase 1),
then the following statements are true with a probability $ 1- o(n^{-4})$.
\begin {enumerate}
\item For $0 < |U_t|< 1/p$, $ (d / 16) |U_t|
      < |U_{t+1}| < (2d) |U_t|$.
\item For $\log^3 n < |U_t| < 1/(p \log n)$,
$\left (1 - 3 / \log n \right ) d |U_t| < |U_{t+1}| < \left (1 + 1 / \log n \right ) d |U_t|$.
\end {enumerate}
\end {lemma}
\begin {proof}
We show this result by bounding the expected number of informed
nodes in each round and then using Chernoff bounds. For a detailed
proof see Appendix \ref {pf:phase1-1}.
\end {proof}

Now, we are ready to show the following concentration result for $|U_{T+1}|$,
the number of active nodes after Phase 1.
\begin {lemma} \label {le:phase1-2}
Let $c_1 = 16^{-4} 4^{-3}$, and $c_2 = 16e$.
After Phase 1 we have with a probability of $1-o(n^{-3})$
$$ c_1 d^T\le |U_{T+1}|\le c_2 d^T.$$
\end {lemma}
\begin {proof}
By Observation \ref {obv:relation}(4), the random experiments
performed in different rounds are independent from each other.
Hence, we can repeatedly use Lemma \ref{le:phase1-1} to bound
$|U_{T+1}|$.

\paragraph {Case 1: $p \ge n^{-4/5}$.}
Since $d = n p \ge n^{1/5}$, $T = \lfloor \log n/\log d  \rfloor \le
4$.  Using Lemma \ref {le:phase1-1}$(1)$ for $T$ rounds, we get
$\left (d/16 \right )^T \le |U_{T+1}| \le (2d)^T $ with a
probability $1 - o(n^{-3})$. To show that we can use Lemma \ref
{le:phase1-1}$(1)$ for Round $i \in [1,T]$, we note that $|U_i| \le
(2d)^{T-1} \le 8 d^{T-1} < 1/p$ since $T \le 4$ and $d \ge \delta
\log n / n$. The lemma now follows from the choices of $c_1$ and
$c_2$.

\paragraph {Case 2: $ n^{-4/5} >p > \delta \log n / n$.}
In this case we have $T = \lfloor\log n/\log d \rfloor \ge 5$. Using
Lemma \ref {le:phase1-1}$(1)$ for three rounds, we get $|U_4| \ge
\left ( d/16 \right ) ^3 > \log ^3 n$ w.h.p since $d = np > \delta
\log n$. Again, we can use Lemma \ref {le:phase1-1}$(1)$ for the
first three rounds. After three rounds, the condition of Lemma \ref
{le:phase1-1}$(2)$ is  w.h.p. fulfilled. In the following we show
that $|U_i|$ does not increase too fast such that we are allowed to
use Lemma \ref {le:phase1-1}$(2)$ for Round $4 \le i \le T-1$,
i.~e.~  $ \log^3 n < |U_i| < 1/(p \log n)$. For the first
inequality, note that $|U_i|$ does not decrease for large
values of $i$ (Lemma \ref {le:phase1-1}$(1)$), w.h.p. For the second
inequality we use Lemma \ref {le:phase1-1}$(1)$ for the first three 
rounds and then Lemma
\ref {le:phase1-1}$(2)$ for the remaining $i - 4$ rounds, we get
$$
|U_{i}| < (2d) ^3 \left (1 + 1 / \log n \right )^{i-4} d^{i-4} < 8
\left ( 1+1 / \log n \right )^{\log n} d^{i-1} < (8e) d^{T-2} < 1 /
(p \log n).
$$
The first inequality uses the fact that $ i < T = \lfloor \log
n/\log d \rfloor \le \log n$. The second inequality uses that
$\forall 0 < x < 1, (1 + x)^{1/x} < e$ and $i \le T-1$. The last
inequality holds because $d^{T-1} < 1/p$ by definition of $T$ and $
d = np > \delta \log n$. This shows that we can use Lemma \ref
{le:phase1-1}$(2)$ for Round $4 \le i \le T-1$. Similarly, we get
$$|U_T| < (2d) ^3 \left (1 + 1 / \log n \right )^{T-4} d^{T-4}
< 8 \left ( 1+1 / \log n \right )^{\log n} d^{T-1} < (8e) d^{T-1} <
1/p, $$ the last inequality holds by $ T = \lfloor \log n/\log d
\rfloor$. This shows that we can use Lemma \ref {le:phase1-1}$(1)$
for Round $T$.

Now we are ready to bound $|U_{T+1}|$. We use Lemma
\ref {le:phase1-1}$(1)$ for three rounds, Lemma \ref
{le:phase1-1}$(2)$ for the next $T - 4$ rounds, and then Lemma \ref
{le:phase1-1}$(1)$ once again. Now we applying the union bound and get with a probability $1 -o(n^{-3})$

$$
\left (d /16 \right )^3 \cdot \left (d \left (1 - 3 / \log n \right
) \right ) ^{T-4} \cdot (d /16)\le |U_{T+1}|
    \le {(2d)^3} \cdot \left ( d \left (1 + 1 / \log n \right ) \right ) ^{T-4} \cdot (2d).
$$
 Since $T \le \log n$, and $\forall 0
\le x \le 1/2, (1 - x)^{1/x} >  1/4$, we have
$$
    \left (d /16 \right )^4 \left ( d \left (1 - 3 / \log n \right ) \right ) ^{T-4}  > \left ({1 / 16} \right )^4 \left ( 1-3 / \log n \right )^{\log n} d^{T}
    > (16^{-4} 4^{-3}) \cdot  d^{T}.
$$
Similarly, we get
$${(2d)^3} \left ( d \left (1 + 1 / \log n \right ) \right )^{T-4} {(2d)}
     < 2^4 \left ( 1+1 / \log n \right )^{\log n} d^{T}
    < (16e) \cdot d^{T}.
$$
This shows that with a probability $1 - o(n^{-3})$ we have
$$
(16^{-4} 4^{-3}) \cdot d^T\le |U_{T+1}| \le (16e) \cdot  d^{T}.
$$
\end {proof}

\subsection {Analysis of Phase 2}

Next we show a result for Phase 2.
If $n^{-2/5} > p > \delta \log n / n$ for a sufficiently large constant $\delta$,
Lemma \ref {le:phase2} shows that after Phase 2 the number of active nodes
is $\Theta (n)$, w.h.p. For the rest case we do not need Phase 2.

\begin {lemma} \label {le:phase2}
Let $c = c_1 4^{-2c_2}/8$. If $n^{-2/5}>p>\delta \log n/n$ for a
sufficiently large constant $\delta$, after Phase 2 (Round $T+1$) we
have with a probability of $1-o(n^{-3})$, $|U_{T+2}| > c \ n$.
\end {lemma}
\begin {proof}
Phase 2 only consists of Round $T+1$ in which every active node
transmits with probability $1/(d^T p)$. We first prove bounds for
$|Q_{T+1}|$. By Lemma \ref {le:phase1-2},
$$
c_2/p > E[|Q_{T+1}|] = |U_{T+1} | \cdot 1/(d^T p) > c_1/p.
$$
Using Chernoff bounds we get
\begin{equation} \label{eq:QT+1}
\Pr [   c_1/2p \le |Q_{T+1}| \le 2c_2 / p  ]
    > 1 - 2e^{-E[|Q_{T+1}|] /4} > 1 - 2e^{- (c_1/p) /4} = 1 - o(n^{-3}).
\end{equation}

Now we fix an arbitrary but not informed node $v$. We show the
probability to inform $v$ in Phase 2 is constant. In order to inform
$v$, $v$ must be connect to exactly one node in $Q_{T+1}$. Hence,
using Equation \ref{eq:QT+1} together with the fact that $\forall
0<x<1/2, (1 - x)^{1/x} > 1/4$, we get
$$
\Pr [v \mbox { is informed}]    = |Q_{T+1}| p (1-p)^{|Q_{T+1}| - 1}
\ge |Q_{T+1}| p (1-p)^{-2c_2/p} > c_1 4^{-2c_2} /2.
$$
Next we show that $N_{T+1} \ge n/2$, w.h.p. First note that we can
assume that $|U_{T+1}| < n/4$. Otherwise, the lemma is already
fulfilled by Observation \ref {obv:relation}$(3)$ and Equation \ref
{eq:QT+1}. This holds since $|U_{T+2}| \ge |U_{T+1}| - |Q_{T+1}| \ge
n/4 - 2c_2/p > n/8$ ($p \ge \delta \log n / n$). Now, using
Observation \ref {obv:relation}$(2)$,
$$
N_{T+1} = n - \left (\sum_{i=1}^{T} {|Q_i|} + |U_{T+1}| \right )
> n - T |U_{T}| - |U_{T+1}|
> n - \log n /p - n/4 > n/2,
$$
with a probability $ 1 - o(n^{-3})$. The first equation follows
since $\forall 1 \le i \le T, Q_i = U_i$ and by Lemma
\ref{le:phase1-1}, $|U_1| < |U_2| < \ldots  <|U_{T}|$. The second
inequality holds since $|U_T| < 1/p$, $T \le \log n$ and $|U_{T+1}|
< n/4$. The third inequality follows since  $p > \delta \log n / n$
for a sufficiently large constant $\delta$.

Next we estimate the expected number of active nodes at the end of
Phase 2.
$$
E[|U_{T + 2}|] = N_{T+1} \Pr [v \mbox { is informed}] \ge (c_1
4^{-2c_2} /4) n.
$$
Note that the events that different not informed nodes are connected
to exactly one node in $U_{T+1}$ are independent from each other.
Also, note that, due to Observation \ref{obv:relation}(4), each of
these events is evaluated only once. Using Chernoff bounds we get
$$\Pr [|U_{T+2}| \le (c_1 4^{-2c_2}/8) n] \le
\Pr [ |U_{T+2}| \le 2 E [|U_{T + 2}|] ] \le e^{-E[|U_{T + 2}|]/4} =
o(n^{-3}).$$
\end {proof}

\subsection {Analysis of Phase 3}

Next, we show that after running Phase 3 for $O(\log n)$ rounds,
every node is informed w.h.p. Note that even at the end of Phase 3,
we still have a considerable amount of active nodes because in each
round of Phase 3, only a small number of active nodes will transmit
and become passive afterwards.

\begin {lemma} \label {le:phase3}
After running Phase 3 for $128 \log n/c$ rounds, every node
is informed with a probability of $ 1 - o(n^{-1})$.
\end {lemma}

\begin {proof}
Let $k = 128\log n/c$. Fix some uninformed node $v$ and let $A_t(v)$
be the number of active neighbours of $v$ at the beginning of Step
$t$ of Phase 3. For any $0 \le t \le k$, let $f_t(v)$ be the number
of active neighbours of $v$ that transmitted before Step $t$ of
Phase 3. Note that $A_t(v) = A_0(v) - f_t(v)$. Let $P_t(v)$ be the
probability to inform node $v$ in Step $t$. In the following we
consider two cases for different values of $p$.
\paragraph {Case 1: $n^{-2/5} \ge p > \delta \log n / n$ for a sufficiently large constant $\delta$.}

We first show that $A_0(v) = \Theta (d)$, w.h.p.. Note that $A_0(v)$
is the number of neighbours of $v$ that are activated in Phase 2.
Since the probability that $v$ is connected to any node in $U_{T+2}$
(the set of nodes that are activated in Phase 2) is $p$, $E[A_0(v)]
= |U_{T+2}| p > cnp = c d$ with a probability at least $ 1-
o(n^{-3})$ by Lemma \ref {le:phase2}. Using Chernoff bounds we get,
\begin {equation} \label {eq:phase3-1}
\Pr [A_0(v) < c d / 2] \le \Pr [A_0(v) < E[A_0(v)]/2] \le e^{-E[A_0(v)] (1/2)^2 / 2} = o(n^{-3}).
\end {equation}
The last inequality holds since $E[A_0(v)] > c np$ with $p > \delta
\log n / n$ for a sufficiently large constant $\delta$.
Similarly, we can show that
\begin {equation} \label {eq:phase3-2}
\Pr [A_0(v) \ge  2 d ] = o(n^{-3}).
\end {equation}

Since every active neighbour of $v$ transmits with probability
$1/d$ in each round of Phase 3,
$E[f_t(v)] \le t A_0(v)/d \le A_0(v) / (4e)$ since $t \le k = 128\log n/c$
and $d = np$ with $p > \delta \log n / n$ for a sufficiently
large constant $\delta$. Using $\Pr [{\mathbb B} (n,p) > a np] < (e/a)^{a np}$ we get,
$$
\Pr [f_t(v) > A_0(v)/2] \le (e / (2e))^{A_0(v)/2} = o(n^{-3}).
$$
The last inequality follows since by Equation \ref {eq:phase3-1},
$A_0(v) > c d /2 > 6 \log n$.
Consequently, it follows by Equation \ref {eq:phase3-1} and \ref
{eq:phase3-2} that $ c d / 4 < A_0(v)/2 < A_0(v) - f_t(v) = A_t(v) <
2d$ with a probability at least $1 - o(n^{-3})$. Using $\forall 0< x
< 1/2, (1 - x)^{1/x} > 1/4$ we get with a probability at least $1 -
o(n^{-3})$,
$$P_t(v) = A_t(v) (1/d) ( 1 - 1/d)^{A_t(v) - 1} \ge c/64.$$
Given this, the probability that $v$ is not informed in $k = 128 \log n /c$ steps 
is at most $(1 - c /64)^k =  o(n^{-2})$.


\paragraph {Case 2: $p > n^{-2/5}$.}
In this case $T = \lfloor \log n / \log d \rfloor = 1$ and using Chernoff bounds we can show that
$3d/4 < |U_2| < 3d/2$ with a probability at least $ 1 - o(n^{-3})$.
Next we show that $A_0(v) = \Theta (d p)$ w.h.p.
Since the probability that $v$ is connected to any active node in $U_2$ is $p$,
$E[A_0(v)] = |U_2| p  \ge 3d p/4$ with a probability at least $ 1- o(n^{-3})$. Using Chernoff bounds we get,
$$
\Pr [ A_0(v) < d p/2 ] \le \Pr [A_0(v) < (2/3) \cdot E[A_0(v)]] \le e^{- E[A_0(v)] (1/3)^2 / 2} = o(n^{-3}).
$$
Similarly, we get
$\Pr [ A_0(v) > 2 d p ] = o(n^{-3})$.
%

The rest proof is very similar to Case 1. In particular, we can show
that with a probability at least $1 - o(n^{-3})$, $dp/4 < A_t(v) <
2d p$. Hence, with a probability at least $1 - o(n^{-3})$,
$$P_t(v) = A_t(v) (1/(dp)) ( 1 - 1/(dp))^{A_t(v) - 1} \ge 1/64.$$
Thus, the probability that node $v$ is not informed at Step $k$ of
Phase 3 is $(1 - 1/64)^k =  o(n^{-2})$. Finally our lemma follows
due to the union bound.
\end {proof}

\section {Gossiping in Random Networks} \label {se:rand-gossip}

In this section we analyse a gossiping algorithm specialised on
random networks. Furthermore, note that similar to \cite
{CGR01,LP02,CR06}, we can obtain a gossiping algorithm with running
time $O(n \log n)$ by combining the framework proposed in \cite
{CGR01} and the broadcasting algorithm in Section \ref
{se:rand-broadcast}. However, the following Algorithm \ref
{alg:gossip} has a better running time of $O(d \log n)$, and it uses
$O(\log n)$ transmissions w.h.p..
Similar to \cite{CR06,CGR01}, we assume that nodes can join messages
originated from different nodes together to one large message, and
we also assume that this message can be sent out in a single time
step.  Let $m_t(u)$ be the message that is send out by node $u$ in
Round t. Then $m_1(u)$ is the message originated in $u$.

\begin {algorithm} 
\caption{A gossiping algorithm for the random network ${\mathbb G} (n,p)$.}
\label {alg:gossip}
\begin{algorithmic}[1]
    \FOR {round $r = 0$ to $128 d \log n$}
        \STATE Every node transmits with probability $1/d$.
        \STATE Every node $u$ joins  $m_{r}(u)$ and any incoming messages to $m_{r+1}(u)$.
     \ENDFOR
\end{algorithmic}
\end{algorithm}


Note that $d=np$ is the average node degree, and diameter $D = (1 +
o(1))(\log n / \log d) < \log n$. Also, note that here
nodes do not become passive after
transmitting once (as it was the case in our broadcasting algorithm
in Section \ref {se:rand-broadcast}). It is easy to see that the
algorithm can be transformed into a dynamic gossiping algorithm. All
that has to be done is to provide every message  with a time stamp
(generation time), and to delete old messages out of the $m_t(i)$
messages.

\begin {theorem} \label {thm:gossip}
Assume $p > \delta \log n /n $ for a sufficiently large constant
$\delta$. Then, with a probability $1 - o(n^{-1})$, 
Algorithm \ref{alg:gossip} completes gossiping in $O(d \log n)$, and every nodes
performs $O(\log n)$ transmissions w.h.p..
\end {theorem}

\begin {proof}
First we bound the gossiping time. Let $u,v$ ($u \ne v$) be an
arbitrary pair of nodes. Let $T$ be the time to send the
gossiping message $m_1(u)$ from $u$ to $v$. Next, we show
that $T$ is w.h.p. at most $128 d\log n$. Fix an arbitrary shortest
path $u=u_1, \ldots u_{L+1}=v$ of length $L\le D$ from $u$ to $v$.
Let $T_i$ be the random variable representing the number of rounds
that it takes node $u_i$ to forward the first message containing
$m_1(u)$ from $u_i$ to $u_{i+1}$. Since $u$ starts to submit its own
message immediately in Round 1, and every node $w$ who receives a
broadcast message in Step $r$ joins the message to its message
$m_{r+1}(w)$, $v$ will get $m_1(u)$ in Step $T \le
\sum_{i=1}^L{T_i}$. It is easy to see that the random variables
$T_1, \ldots, T_L$ are independent from each other. To bound $T$, we
first prove a result which is similar to Lemma $3.4$ in \cite
{CR06}.

\begin {lemma}\label{geoproof}
Let $Y_1, \ldots, Y_L$ be a sequence of geometrically distributed
random variables with parameter $1/(16d)$, i.~e.~, $\forall 1 \le i
\le L, k \ge 1, \Pr [Y_i = k] = 1/(16d) ( 1 - 1/(16d))^{k-1}$. Then
$T \prec \footnote {We say a random variable $A$ is {\em
stochastically dominated} by another random variable $B$, writing $A
\prec B$, if $\forall k \in {\mathbb R}$, $\Pr [A > k] \le \Pr [B >
k]$. } \sum_{i=1}^L{Y_i}$ with a probability at least $1-o(n^{-3})$.
\end {lemma}

\begin {proof}
The proof can be found in Section \ref{geo} of the appendix.
\end{proof}

\section {Broadcasting in General Network} \label {se:arbitrary}

In this section we consider broadcasting on arbitrary networks with
diameter $D$. Czumaj and Rytter (\cite{CR06}) propose a randomised
algorithm with $ O(\log^2 n + D \log (n/D))$ broadcasting time.
Their algorithm can easily be transformed into an algorithm with the
same runtime and an expected number of $\Omega(\log ^2 n)$
transmissions per node. The only modification necessary is to stop
nodes from transmitting after a certain number of rounds (counting
onwards from the round they got the message for the first time). In
Czumaj and Rytter's algorithm, each active node transmits with
probability of $\Theta (1/ \log (n/D))$ per round. It informs an
arbitrary neighbour $u$ (i.~e.~ it transmits the message {\em and}
is the only neighbour of $u$ that transmits in that round) with a
probability of $\Omega (1 / (\log (n/D) \log n))$ per round. Hence,
to get a high probability bound, every node has to try to send a
message for $O(\log ^2 n \log (n/D))$ rounds. Since an active node
transmits with probability $O(1/ \log (n/D))$, the total expected
number of transmissions is $O(\log ^2 n)$ per node. Similarly, the
algorithm of \cite{CR06} for unknown diameter can be transformed
into an algorithm with an expected number of $O(\log^2 n)$ messages
per node.

Unfortunately, in general the expected number of $O(\log^2 n)$
transmissions per node can not be improved without increasing the
broadcasting time (see Corollary \ref{cor:lower-bound}). Under the
assumption that the network diameter $D$ is known in advance, we
propose a new randomised oblivious algorithm with broadcasting time
$O(D \log (n /D) + \log^2 n)$ that uses only an expected number of
$O(\log ^2 n / \log (n /D))$ transmissions per node (see Section
\ref{se:ub-arbitrary}). Note that our algorithm achieves the same
broadcasting time as the algorithm in (\cite {CR06}). In Section
\ref {se:lb-arbitrary}, we prove a matching lower bound on the
number of transmissions (Theorem \ref {thm:lb2}) which indicates
that our proposed algorithm is optimal in terms of the number of
transmissions. In Theorem \ref {thm:tradeoff} we show a trade-off
between broadcasting time and number of transmissions.

\subsection {Upper Bound for Broadcasting} \label {se:ub-arbitrary}
In this section we show that, if the graph diameter $D$ is known in
advance, the number of transmissions can be reduced from $O(\log ^2
n)$ to $O(\log ^2 n / \log (n/D))$. The improvement is due to a new
random distribution which is defined in Figure \ref {fig:prob}.
Let $\lambda = \log (n/D)$. The  distribution we use to
generate the randomised sequence is denoted by $\alpha$, and the
distribution used in Section 4.1 of \cite {CR06} is denoted by
$\alpha'$. See Figure \ref {fig:prob} for a comparison of the two distributions. Note that
$\forall 1 \le k \le \log n$, $1/(2 \log n) \le \alpha_k \le 1/(4
\lambda)$ and $\alpha_k \ge \alpha'_k /2 $.

\begin {figure}[ht]
\begin {minipage}[ht]{0.36\linewidth}
\begin {displaymath}
\alpha_k = \left \lbrace
    \begin {array}{ll}
    \frac{1}{4 \lambda}  \\
    \max \{ \frac{1}{2 \log n}, \frac{1}{2 \lambda} 2 ^{-(k - \lambda)} \}  \\
    \frac{1}{2 \log n}   \\
    1 - \sum_{i = 1}^{\log n} {\alpha_i}
    \end {array}
    \right.
\end {displaymath}
\end {minipage}
\begin {minipage}[ht]{0.64\linewidth}
\begin {displaymath}
\alpha'_k = \left \lbrace
    \begin {array}{ll}
    \frac{1}{2 \lambda}  & 1 \le k \le \lambda  \\
    \frac{1}{2 \lambda} 2 ^{-(k - \lambda)}  &\lambda < k \le \min \{\lambda + \log \log n,  \log n \}, \\
    \frac{1}{2 \lambda \log n} & \log \log n + \lambda < k \le \log n, \\
    1 - \sum_{i = 1}^{\log n} {\alpha'_i} & \mbox { for } k = 0.
    \end {array}
    \right.
\end {displaymath}
\end {minipage}
\caption {Comparison of our distribution (left) vs. the distribution
in \cite {CR06} (right) \label {fig:prob} }
\end {figure}

\begin{algorithm}
\caption{An energy efficient broadcasting algorithm for arbitrary
network with diameter $D$} 
\label{alg:arbitrary}
\medskip
\begin{algorithmic}[1]
\STATE {Choose a randomised sequence $\Gamma = <I_1, I_2, \ldots, >$ such that \\
$\Pr [I_r = k] = \alpha_k, \forall r \in {\mathbb N}, \forall k \in \{0,1, \ldots, \log n \}$.}
\STATE {The status of the source is set to {\em active}.}
\FOR {$r = 1$ to $T$ every active node $u$}
        \IF {$r \le t_u + \beta \log^2 n$ ($\beta$ is a constant)}
            \STATE {$u$ transmits with probability $2^{-I_r}$.}
        \ELSE
            \STATE {$u$ becomes {\em passive}.}
        \ENDIF
    \IF {$u$ receives the message {\em for the first time}}
    \STATE the status of $u$ is set to {\em active}.
    \ENDIF
    \ENDFOR
\end{algorithmic}
\end{algorithm}

We prove the following theorem. Note that the broadcasting time is
optimal according to the lower bounds shown in \cite {KM98} and
\cite {LP02}.

\begin {theorem} \label {thm:upper-bound}
Algorithm \ref {alg:arbitrary} completes broadcasting in $O(D \log
(n/D) + \log ^2 n)$ rounds with probability at least $1 - n^{-1}$.
The expected number of messages per node is $O(\log^2 n / \log
(n/D))$.
\end {theorem}

\begin{sketch}
Each node is active for $O(\log ^2 n)$ rounds. In every round, an active node transmits with a
probability of $O(1/ \log (n/D))$. Hence, the expected total number
of transmissions is $O(\log^2 n / \log (n/D))$ per node.

To show that every node receives the broadcast message, fix a round
$r$, an arbitrary active node $v$ and one of its neighbors $w$.
Assume $w$ has $m \ge 1$ active neighbors in Round $r$ and let $1
\le k \le \log n$ such that $w/2 < 2^k < w$. If every active
neighbor of $w$ sends with probability $2^{-k}$ (i.~e.~ $I_r=k$),
$w$ is informed with probability at least $0.1$  according to Lemma
3.2 in \cite {CR06}.
 For any $1 \le x \le \log n$, $\alpha_x \ge 1/(2 \log n)$,
 $I_r = k$ with probability at least $1/(2 \log n)$. Hence,
 the probability to inform $w$ is at least $ 1/(20\log n)$ per round.
Using Chernoff bounds we can show that $v$ can successfully inform
all its neighbours, w.h.p..

To bound the broadcasting time, we compare the runtime of our
algorithm with the runtime of the algorithm for shallow networks in
\cite {CR06}. Any send probability that is chosen by the
algorithm in \cite {CR06} is chosen with at least half the
probability by our algorithm. Thus, we can use a proof that is
similar to the proof of Theorem $2$ in \cite {CR06} to show our
result.
\end {sketch}

Finally, we demonstrate that there is a tradeoff between the
expected number of transmissions and the broadcasting time.

\begin {theorem} \label{thm:tradeoff}
Let $\log (n/D) \le \lambda \le \log n$. Algorithm \ref
{alg:arbitrary} finishes broadcasting in $ O(D \lambda + \log ^2 n)
$ rounds w.h.p.. The expected number of transmissions is $O(\log^2 n
/ \lambda)$ per node.
\end {theorem}

\begin {sketch}
Every node is active for $O(\log ^2 n)$ rounds.
Moreover, the expected number of transmissions an active node performs in every round is $O(1/\lambda)$. Hence the expected total number of transmissions is $O(\log^2 n / \lambda)$ per node.
Since for all $1 \le k \le \log n$, $\alpha_k \ge 1/(2 \log n)$,
 we can show (similar to the proof of Theorem \ref {thm:upper-bound}) that
every node receives the broadcasting message w.h.p..


It remains to bound the broadcasting time. Our proof is
similar to the proof of Theorem $2$ in \cite {CR06}. We first fix
some shortest path $v_0, \ldots, v_L$ of length $L \le D$ from the
source to an arbitrary node. Then, we partition all nodes into $L$
disjoint layers with respect to that path. We assign a node $u$ to
layer $i, 1 \le i \le L$, if node $v_i$ is the highest ranked node
on the path that $u$ has an edge to. In the following, a layer is
called {\em small}, if its size is smaller than $2^\lambda$,
otherwise it is called {\em large}.

For an arbitrary small layer, since $\forall 1 \le k \le \lambda$,
$\alpha_k \ge 1/(4\lambda)$, use a similar argument as in Theorem
\ref {thm:upper-bound}, we get that the probability to inform some
node in the next layer is at least $1/(40 \lambda)$. Hence the
expected time spent on any small layer is $O(\lambda)$. Since there
are at most $D$ layers and by applying the concentration bound in
Lemma $3.5$ of \cite {CR06}, we get that the total time spent on all
small layers is $O(D \lambda)$ w.h.p..

For an arbitrary large layer (of size $s 2^\lambda, s > 1$),
since $\forall \lambda < k \le \log n$,
$\alpha_k \ge \frac{1}{2 \lambda} 2^{-(k-\lambda)}$,
similar to Theorem $2$ in \cite {CR06}, we can show that
the probability to inform some node in the next layer is $\Omega (1/(s\lambda))$.
Hence, the expected time spent on a large layer is $O(s \lambda)$.
Consequently, the total expected time spent on all large layers is
$O(\lambda n / 2^\lambda) = O(D \lambda)$ since $2^\lambda \ge n/D$.
Applying Lemma $3.5$ in \cite {CR06} once again, we obtain the high probability bound.
\end {sketch}



\subsection {Lower Bound on the Transmission Number} \label {se:lb-arbitrary}
In this section we show  two lower bounds for oblivious broadcasting
algorithms.  Observation \ref
{obv:lb1}, shows a lower bound on the expected number of
transmissions for any randomised oblivious (
every node uses the same algorithm) broadcasting algorithm.
We call a probability distribution  {\em time-invariant} if it does
not depend on the time $t$. Theorem \ref {thm:lb2} shows a lower
bound on the expected number of transmissions of any optimal
randomised oblivious algorithm using a time-invariant distribution.

\begin {observation} \label{obv:lb1} Let $A$ be an oblivious broadcast
  algorithm. Then,
  for every $n$ there exists a network with $O(n)$ nodes such that $A$
  needs at least $n \log n /2$ transmissions to complete broadcasting
  with a probability of at least $1 -n^{-1}$.
\end {observation}
\begin {proof}
The proof can be found in Section \ref{observ} of the appendix.
\end {proof}


Next, we show a matching lower bound result on the number of transmissions. This result holds for a group of
randomised oblivious algorithms with optimal (i.~e.~ $O (D \log
(n/D))$) broadcasting time (e.~g~. the algorithm in \cite {CR06}).

\begin {theorem} \label{thm:lb2} Let $D > 1$, let $c,i$ be constants,
  and fix an arbitrary $n=2^i$.  Let $A$ be an
  oblivious broadcast algorithm using a time-invariant probability
  distribution $\alpha$.  For every $n > 0$, there is a network with
  $O(n)$ nodes and diameter $D$, such that $A$ requires an expected
  number of at least
$$\log^2 n /(\max\{4c, 8\} \log (n/D))$$
transmissions per node in order
to finish broadcasting in $ c D \log (n/D)$
rounds with probability at least $1- n^{-1}$.

\end {theorem}

\begin {proof}
We can assume that $D > 4 \log n$,  otherwise this result can be
obtained directly from Observation \ref {obv:lb1} since $\log (n/D)
> \log n/2$. We construct a layered network (See Figure \ref
{fig:lowerbound}) consisting of two subgraphs $G_1$ and $G_2$. $G_1$
has $\log n$ layers, namely $S_1, \ldots, S_{\log n}$, where $S_i, 1
\le i \le \log n$ is a star consisting of one center node $c_i$ and
$2^i$ leaf nodes. Every leaf node in $S_i$ has an edge to the center
$c_{i+1}$ of $S_{i+1}$, for $1 \le i \le \log n - 1$. $G_2 = v_0,
\ldots, v_L$ is a path of length $L = D - 2 \log n$. To connect
$G_1$ and $G_2$, we connect every node of the star $S_{\log n}$ to
the first node of $G_2$, also denoted as $c_{\log n + 1}$. Note that
our network has $\sum_{i=1}^{\log n}{(2^i +1)} + D - 2 \log n + 1
\le 2n + D$ nodes and diameter $D$.

 \begin{figure}[ht]
 \centerline{
   {\scalebox{0.65}{\includegraphics{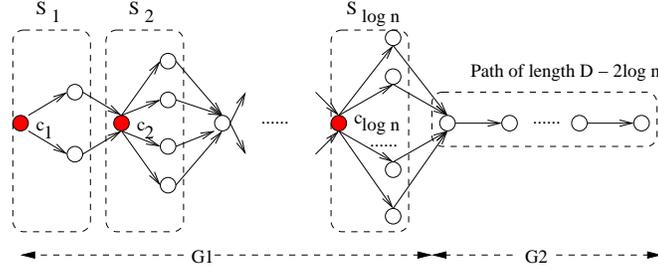}}}}
 \label {fig:lowerbound}
 \caption{The network used in the lowerbound proof}
 \end{figure}

We assume that $c_1$ is the originator of the broadcast. The purpose
of $G_1$ is to show that every informed node in $G$ must be active
for at least $\ln ^2 n$ rounds in order to complete broadcasting
with probability $1 - n^{-1}$. More specifically, no matter what
$\alpha$ is, there is always a star $S_i$ such that the probability
to inform $c_{i+1}$ is at most $1/\ln n$. Since our distribution is
time invariant and every node does not know which star it belongs
to, every node in the network needs to be active for at least $\ln^2
n$ rounds. Let $\mu$ be the mean of distribution $\alpha$ and
$ran(\alpha)$ be the set of outcomes of $\alpha$.
Next, we use $G_2$ to argue that in order to finish broadcasting in
$cD \log (n/D)$ rounds, $\mu$, mean of $\alpha$, must be at least
$1/(2c \log (n/D))$. Hence, the total expected number of
transmissions per node is at least $\ln ^2 n (1/(2c \log (n/D)) >
\log ^2 n / (4c \log (n/D))$.

Let $A_i$ be the event that $c_{i+1}$ is informed in Round $t_i$
under the condition that every leaf node of $S_i$ is active (note
that they are always activated at the same time). Let $Q_{t_i}$ be
the random variable that represents the probability chosen at Round
$t_i$. Note that $Q_{t_i}$ has distribution $\alpha$. For any $q \in
ran(\alpha)$, let $\Pr[A_i|Q_{t_i} = q]$ be the probability to
inform $c_{i+1}$ if $Q_{t_i}=q$. Since $c_{i+1}$ is informed if
exactly one of the $2^i$ leaf nodes of $S_i$ transmits we get

\begin {equation} \label {eq:prob}
\Pr[A_i|Q_{t_i} = q] = 2^i q ( 1 - q)^{2^i-1}
< 2^i q e^{-(2^i-1)q}.
\end {equation}
Observe that $\Pr[A_i] = \sum_{q \in ran(\alpha)}{(\Pr [ Q_{t_i} =
q] \Pr[A_i | Q_{t_i} = q])}$. We get,
\begin {eqnarray*}
\sum_{i=1}^{\log n} {\Pr[A_i]}  &=&
\sum_{i=1}^{\log n} \sum_{q \in ran(\alpha)} {\left (\Pr [ Q_{t_i} = q] \Pr[A_i | Q_{t_i} = q] \right )} \\
                                &=&
\sum_{q \in ran(\alpha)} {\left (\Pr [ Q_{t_i} = q] \sum_{i=1}^{\log
n} {\Pr[A_i | Q_{t_i} = q]} \right )}
                \le \left (\sum_{q \in ran(\alpha)} {\Pr [ Q_{t_i} = q]} \right ) \frac{1}{\ln 2}
                 = \frac{1}{\ln 2}.
\end {eqnarray*}

For the third inequality, we use Equation \ref {eq:prob} and $
\forall 0 \le q \le 1, \int_{1}^{\infty} 2^iq e^{-(2^i-1)q} di =
1/(e^{q} \ln 2) \le 1/\ln 2$. Consequently,
$$
\min_{i} {\Pr[A_i]} \le \left ( \sum_{i=1}^{\log n} {\Pr[A_i]}
\right ) / \log n \le \frac{1}{\ln 2 \log n} = \frac{1}{\ln n}.
$$
Let $i^* = \arg\!\min_{i} {\Pr[A_i]}$. Consequently, in order to
complete broadcasting with probability at least $ 1 - n^{-1}$, every
leaf node of $S_{i^*}$ must be active for at least $\ln^2 n$ rounds.

In the following we show $\mu \ge 1/(2c \log (n/D))$ using $G_2$.
First note that $L = D - 2 \log n > D/2$ since $D \ge 4 \log n$. For
any $0 \le i \le L-1$, let $T_i$ be the number of rounds that $v_i$
is the highest ranked node on the path that is informed. Note that
$T_i$ is geometrically distributed with probability $\mu$, we have
$E [\sum_{i=0}^{L-1}{T_i}] = L \cdot E[T_i] = L/\mu$. Hence, in order to
inform $v_L$ within $c D \log (n/D)$ steps (even expectedly), we
need $\mu \ge 1/(2c \log (n/D))$ since $L > D/2$.

We have shown that every node in the network needs to be active for
$\ln^2 n$ rounds while in each round, the expected number of
transmissions it performs is at least $1/(2c \log (n/D))$. Hence,
the total expected number of transmissions per node is $(\ln ^2 n)
(1 /(2c  \log (n/D))) > \log^2 n / (4c \log (n/D))$.
\end {proof}

Setting $D = n$ in the network constructed above, we immediately get
the following corollary.

\begin {corollary} \label{cor:lower-bound}
There exists a network with $O(n)$ nodes such that any randomised
oblivious broadcasting algorithm that finishes broadcasting in $c n$
rounds with probability at least $ 1 - n^{-1}$ requires an expected
number of $\Omega (\log^2 n)$  transmissions.
\end {corollary}

\newpage

\appendix

\noindent {\Large \bf Appendix}

\section {Chernoff Bounds} \label {se:Chernoff}
Here we present a version of Chernoff bounds, which can be found, for example, in, \cite {MU05}.
\begin {lemma}
Let $X_1, \ldots X_n$ be independent Bernoulli random variables
 and let $X = \sum_{i=1}^{n}{X_i}$ and $\mu = E[X]$. Then we have,
\begin {enumerate} \itemsep 0pt
\item $\Pr \left [X < (1 - \epsilon) \mu \right ] < e^{-\mu \epsilon^2/2}$, for $0 \le \epsilon \le 1$.
\item $\Pr \left [X > (1 + \epsilon) \mu \right ] < e^{-\mu \epsilon^2/3}$, for $\epsilon > 0$.
\item $\Pr \left [ |X - \mu| \le \epsilon \mu \right ] > 1 - 2 e^{-\mu \epsilon^2/3}$, for $0 \le \epsilon \le 1$.
\end {enumerate}
\end {lemma}

\section {Proof of Lemma  \ref {le:phase1-1}} \label {pf:phase1-1}
\begin {proof}
We consider two cases of different values of $p$.
If $p > 1/2$, we have $T=1$ and every node will have expectedly $(n-1)/2$ neighbours.
The result now follows from a simple application of Chernoff bounds.
If $p \le 1/2$, we fix an arbitrary node $u$ and a round $t=1$ in Phase 1.
First we bound $q$, the probability that $u$ is informed in Round $t$,
i.~e. $u$ is connected to {\em exactly} one node in $U_t$.
\begin{equation}\label{equq}
q    = |U_t| p ( 1- p)^{|U_t|-1}
> p |U_t| ( 1- p)^{1/p} \ge p|U_t|/4.
\end{equation}
Here, the first inequality uses the condition $|U_t|< 1/p$. To see the second one, note that $\forall 0< p < 1/2,
( 1 - p)^{1/p} > 1/4$.
Next, we show $N_t$, the number of not informed nodes
at time $t$, is larger than $n/2$.
By Observation \ref{obv:relation}$(2)$,
\begin{equation}\label{nt}
N_t = n - \left ( \sum_{i=1}^{t-1} {|Q_i|} + |U_t| \right )
  > n - t|U_t|
  > n - (\log n) (1/p)
    > n/2.
\end{equation}
Here, the first inequality is true by Observation \ref{obv:relation}$(1)$ and $|U_1| < |U_2| < \ldots <|U_t|$.
The second one uses the condition $|U_t| < 1/p$
and $t \le T = \lfloor \log n/\log d \rfloor \le \log n$.
The third inequality uses  $p > \delta \log n / n$. Hence,
$$
E[|U_{t+1}|] = N_t q > (n/2) \cdot q \ge (n/2) \cdot p|U_t|/4 = d|U_t|/8,
$$
since $N_t > n/2$ and $d = n p$.
Note that the events to be connected to exactly one node in $U_t$ are
independent for different not informed nodes. Also, note that each event
is only evaluated once due to Observation \ref{obv:relation}(4).
Using Chernoff bounds we get
$$
\Pr [U_{t+1}| \le d|U_t|/16 ]
    \le  \Pr [|U_{t+1}| \le E[|U_{t+1}|]/2]
    \le e^{-d |U_t|/64} = o(n^{-4}).
$$
The last inequality uses $d=np$ with $p = \delta \log n /n$ for a sufficiently large constant $\delta$.
Consequently $|U_{t+1}| / |U_t| > d/16$ with a probability $1 - o(n^{-4})$.
Using a similar approach, we can prove that $|U_{t+1}| / |U_t| < 2d$ with a probability
$1 - o(n^{-4})$. This finished the proof of Part 1 of the lemma.

\bigskip

To prove part 2 we first need a tighter bound on $q$. By Equation \ref {equq},
$$
 q  = |U_t| p \cdot (1 - p)^{|U_t|-1} > ( 1 - p|U_t|) \cdot p |U_t| > \left ( 1 - 1 / \log n \right ) \cdot p|U_t|
$$
Next we bound $N_t$. Using Equation \ref{nt} with $|U_t| < 1/(p \log n)$ and
$t \le T = \lfloor\log n/\log d \rfloor \le \log n$ we get
$$
N_t = n - \left (\sum_{i=1}^{t-1}{|Q_i|} + |U_t| \right ) >
n - t |U_t| > n - 1/p > n \left (1 - 1 / \log n \right ).
$$
Now, we obtain the following lower bound for $E[|U_{t+1}|]$,
$$
E[|U_{t+1}|]= N_t q > \left ( 1 - 1 / \log n \right )^2 \cdot d |U_t| > \left (1 - 2 /\log n \right ) \cdot d |U_t|.
$$
For an upper bound on $E[|U_{t+1}|]$ we use $N_t < n$ and $q \le  p |U_t|$ to get
$$
E[|U_{t+1}|] = N_t q < n p |U_t| = d |U_t|.
$$
Using Chernoff bounds together with the assumption that $|U_t| > \log ^3 n$, we get
$$
\Pr \left [\left (1 - 3 / \log n \right ) d |U_t| < |U_{t+1}| < \left (1+ 1 / \log n \right ) d |U_t| \right ] > 1 - 2e^{-\frac{E[|U_{t+1}|]}{4 \log^2 n}} = 1 - o\left ( n^{-4} \right ).
$$
\end {proof}

\section {Proof of Lemma \ref {geoproof}} \label {geo}

The proof is similar to the proof of Lemma $3.4$ in \cite {CR06}. All that we
have to do is to bound the probability $q$ that a node successfully
sends a message to a fixed neighbour. The expected degree of every
node is $d$ and using Chernoff bounds we can show the degree of
every node is at most $2d$ with a probability $1 - o(n^{-5})$.
Hence, with a probability $1- o(n^{-5})$, we have
$$q_r \ge (1/d) ( 1 - 1/d)^{2d-1} \ge  (1/d) \cdot (1 - 1/d)^{2d} \ge (1/d )\cdot (1/4)^2 = 1/(16d).$$
\end {proof}

Now it remains to bound $\Pr [\sum_{i=1}^L{Y_i} \le 128 d \log n]$.
Similar to the proof of Lemma $3.5$ of \cite {CR06}, applying the
standard relation of geometric distribution and binomial
distributions, and using Chernoff bounds on the corresponding
binomial distribution, we get
$$
\Pr \left [\sum_{i=1}^L{Y_i} > 128d \log n \right ]
    \le \Pr \left [{\mathbb B} (128d \log n, 1/(16d) ) < L \right ]
    \le e^{-  (7/8)^2 \cdot 8\log n / 2} < e^{- 3 \log n} = o(n^{-3}).
$$
The third inequality holds since $L \le  D < \log n$. The bound on
the gossiping time follows by the union bound and the fact that
there are in total $n(n-1)$ source-destination pairs.

Next we bound the number of transmissions. Let $v$ be an arbitrary
node and denote $Z_v$ to be the number of transmissions performed by
$v$. Note that $ E[Z_v] = 128 \log n$ since in each round, every
node transmits with probability $1/d$ and our algorithm has in total
$128 d \log n$ rounds. Using Chernoff bounds we get that $Z_v \le
256 \log n$ with probability $1 - o(n^{-2})$. By the union bound, we
get with a probability $ 1 - o(n^{-1})$, none of the nodes performs
more than $256 \log n$ transmissions.

\section {Proof of Observation \ref {obv:lb1} \label {observ}}

We construct a network with $3n+1$ nodes. $s$ is the node initiating
the broadcast, and $d_1, \ldots, d_n$ are the destination
nodes. $s$ has an edge to $2n$ intermediate nodes $u_1, \ldots
u_{2n}$. For all $1\le i \le n $, $d_i$ connects to both $u_{2i-1}$
and $u_{2i}$. Let us assume that $s$ informs $u_1,\ldots, u_{2n}$ in
Round $t_1$. Now fix some arbitrary $T > t_1$. In Round $t_1+1 \le r
\le T$, let $q_r$ be the send probability used by the algorithm. For
all $1 \le i \le n$, the probability  to inform node $d_i$ in Round
$r$ is $2q_r ( 1 - q_r)$. Due to symmetry we can assume that $q_r
\le 1/2$, resulting in $ (1 - q_r)^{1/q_i} \ge 1/4$. Hence,

\begin {eqnarray*}
\Pr [ d_i \mbox { is not informed before Round $T$}] 
 & =  &\prod_{r = t_1+1}^T {\left (1 - 2q_r ( 1 - q_r) \right )}\\
    > \prod_{r = t_1+1}^T { (1 - q_r)^2} 
    &\ge& \prod_{r = t_1+1}^T {4^{-2 q_r}} = 2^{-4 \sum_{r = t_1+1}^T{q_r}}.
\end {eqnarray*}
Now it is easy to see that, to inform $d_i$ with probability $1 -
n^{-1}$, we need $\sum_{r = t_1+1}^T{q_r} >  \log n/4$. Note that
$\sum_{r = t_1+1}^T{q_r}$ is the expected number of transmissions
that $u_i$ and $v_i$ perform between Round $t_1+1$ and $T$. The
total number of transmissions performed by all $2n$ intermediate
nodes
 is at least $2n \ (\log n /4) = n \log n /2$.



\end{document}